\documentclass[final,5p,times,twocolumn]{elsarticle}

\usepackage{amssymb}
\usepackage{amsmath,amsfonts}
\usepackage{array}
\usepackage{textcomp}
\usepackage{url}
\usepackage{verbatim}
\usepackage{graphicx}
\usepackage[ruled, boxed]{algorithm2e}
\usepackage{booktabs}
\usepackage{hyperref}
\usepackage{multirow}
\usepackage{subfigure}
\usepackage{xcolor}
\usepackage[inline,shortlabels]{enumitem}
\hypersetup{colorlinks=true,linkcolor=black,citecolor=blue}
\graphicspath{{./figures/}}

\journal{Engineering Applications of Artificial Intelligence}

\begin{document}

\begin{frontmatter}



  \title{Mixed-Integer Optimal Control via Reinforcement Learning: A Case Study on Hybrid Electric Vehicle Energy Management}


  \author[a]{Jinming Xu}
  \author[b]{Nasser Lashgarian Azad}
  \author[a]{Yuan Lin\corref{cor}}
  \ead{yuanlin@scut.edu.cn}
  \cortext[cor]{Corresponding author}

  \affiliation[a]{organization={Shien-Ming Wu School of Intelligent Engineering},%Department and Organization
    addressline={South China University of Technology},
    city={Guangzhou},
    postcode={511442},
    state={Guangdong},
    country={China}}
  \affiliation[b]{organization={Department of Systems Design Engineering},
    addressline={University of Waterloo},
    city={Waterloo},
    postcode={N2L 3G1},
    state={Ontario},
    country={Canada}}

  \begin{abstract}
    Many optimal control problems require the simultaneous output of discrete and continuous control variables.
    These problems are usually formulated as mixed-integer optimal control (MIOC) problems,
    which are challenging to solve due to the complexity of the solution space.
    Numerical methods such as branch-and-bound are computationally expensive and undesirable for real-time control.
    This paper proposes a novel hybrid-action reinforcement learning (HARL) algorithm, twin delayed deep deterministic actor-Q (TD3AQ), for MIOC problems.
    TD3AQ combines the advantages of both actor-critic and Q-learning methods, and can handle the discrete and continuous action spaces simultaneously.
    The proposed algorithm is evaluated on a plug-in hybrid electric vehicle (PHEV) energy management problem, where real-time control of the discrete variables, clutch engagement/disengagement and gear shift, and continuous variable, engine torque, is essential to maximize fuel economy while satisfying driving constraints.
    Simulation outcomes demonstrate that TD3AQ achieves control results close to optimality when compared with dynamic programming (DP), with just 4.69\% difference.
    Furthermore, it surpasses the performance of baseline reinforcement learning algorithms.
  \end{abstract}

  %   \centering
  %   \includegraphics[width=0.98\textwidth]{Graphical Abstract.pdf}

  % \begin{highlights}
  %   \item A novel reinforcement learning algorithm for mixed-integer optimal control problems
  %   \item Comparative analysis of different hybrid-action reinforcement learning algorithms for hybrid electric vehicle energy management
  %   \item Near-optimal control with 4.69\% difference from dynamic programming achieved with the proposed algorithm
  % \end{highlights}


\end{frontmatter}


\section{Introduction}\label{sec:introduction}
Mixed-integer optimal control (MIOC) problems commonly arise in various engineering applications, such as assignment problems~\cite{wang2021multi}, control of piecewise-affine (PWA) systems~\cite{sun2022piecewise} and hybrid systems~\cite{salahshoor2013novel}.
These problems are particularly challenging as they require optimization of a system that involves both continuous and discrete control variables.
Traditional numerical optimization methods, such as branch-and-bound and heuristic algorithms, are computationally expensive or rely on system models~\cite{richards2005mixed,fischetti2010heuristics,belotti2013mixed}, making them unpreferred for general real-time control applications. Recent advances in reinforcement learning (RL) have demonstrated promising results in handling complex control problems~\cite{silver2017mastering,vinyals2019grandmaster,degrave2022magnetic}. However, most existing RL algorithms are tailored for either continuous~\cite{lillicrap2015continuous} or discrete~\cite{silver2017mastering} action spaces, and are not directly applicable to MIOC problems. Currently, research on MIOC through RL is still in its nascent stages.
This paper proposes an innovative hybrid-action reinforcement learning (HARL) algorithm for MIOC problems, and applies it to a plug-in hybrid electric vehicle (PHEV) energy management problem to validate its effectiveness.

\subsection{Literature Review}
Various methods have been proposed to solve MIOC problems. However, the unique characteristics of these methods still present certain limitations.

One approach is to formulate MIOC problems as mixed-integer programming (MIP) forms and solve them using the branch-and-bound method~\cite{morrison2016branch}. This method typically involves dividing the problem into smaller sub-problems or branches, and then solving each branch separately. A bounding function is used to determine if each branch can be pruned or eliminated, until the optimal solution is found. While this method can achieve globally optimal solutions, it is computationally expensive and time-consuming.

Another approach is to use dynamic programming (DP), which breaks the problem into smaller sub-problems and solves them recursively~\cite{bellman2010dynamic}.
However, DP cannot handle continuous variables naturally and requires discretization of the continuous variables into a finite set of discrete values~\cite{bertsekas2012dynamic}. A fine-grained discretization can lead to a more accurate solution, but it also increases the computational cost, making it unsuitable for large-scale problems. Additionally, DP is not suitable for real-world optimal control since it requires knowledge of the entire state space in advance.

Heuristic algorithms, while not providing optimality guarantees, can offer feasible and fast solutions to MIOC problems~\cite{fischetti2010heuristics,belotti2013mixed}. These methods use problem-specific knowledge to guide the search for a solution, rather than exhaustively examining every possible solution. They are simple to design and can be applied to a range of combinatorial optimization problems. However, it requires specific implementation and modification to match the problem structure being solved~\cite{belotti2013mixed}.

The development of reinforcement learning (RL) has led to a new paradigm for solving complex control problems~\cite{vinyals2019grandmaster,kaufmann2023champion}.
RL learns a policy that maps the current state of the system to an action that maximizes the expected cumulative reward. With the learned policy, RL can reach a near-optimal solution without requiring any prior knowledge of the environment~\cite{sutton2018reinforcement}.
However, existing RL algorithms are mostly suitable for continuous or discrete action spaces, which cannot be directly applicable to MIOC problems.

Discretizing continuous variables into finite discrete values and applying discrete RL methods, such as Rainbow~\cite{hessel2018rainbow}, is a straightforward approach to tackle MIOC problems.
However, this approach becomes impractical for high-dimensional control problems as it necessitates a substantial number of discrete values to achieve accurate outcomes.

In recent years, a substantial amount of RL research has focused on a special type of MIOC problems called parameterized action Markov decision process (PAMDP)~\cite{masson2016reinforcement}.
In PAMDP, the core idea involves structuring the problem hierarchically, with each discrete action linked to its own continuous parameters.
One approach is to alternate between optimizing the discrete action selection with Q-learning and the parameter selection with a policy search method~\cite{masson2016reinforcement}.
Another approach called parameterized deep deterministic policy gradient (PA-DDPG) is to directly output the continuous parameters and weights for discrete actions from DDPG~\cite{lillicrap2015continuous} and select the discrete action with the highest weight~\cite{hausknecht2015deep}.
Parameterized deep Q-networks (P-DQN) combine the actor-critic architecture and DQN, where the actor network first generates parameters for every discrete action, followed by the Q-network's selection of the action that offers the maximum Q-value~\cite{xiong2018parametrized}.
However, the formulation of P-DQN is flawed because the discrete action values depend on all continuous action parameters, not just those associated with each action. Therefore, Bester et al.~\cite{bester2019multi} proposed a multi-pass DQN (MP-DQN) that used multiple forward passes from the actor network to the Q-network. Each pass involves only one set of action parameters for a discrete action.

In addition to the learning-based framework, Massaroli et al.~\cite{massaroli2020neural} utilized neural ordinary differential equations as state-value networks to provide optimal action parameters for all discrete actions, then obtained the optimal discrete action using a value-based method.
Li et al.~\cite{li2022hyar} constructed the latent representation space of discrete actions and continuous parameters using a conditional variational auto-encoder. The RL agent's output of the action and its parameters are initially fed into the representation model and subsequently exported to the environment.
Overall, these methods may not be best for general MIOC problems as they rely on the parameterized action space.

Alternatively to PAMDP, some methods handle the mixed-integer action space in a unified manner. Jiang et al.~\cite{jiang2019hd3} proposed a distributed dueling framework based on DQN, which added an extra fully-connected layer to generate continuous actions prior to the advantage and value layers. The discrete action was then selected to maximize the Q-value. Fan et al.~\cite{fan2019hybrid} utilized a hybrid actor-critic architecture, two parallel actor networks were used to generate continuous and discrete actions, respectively. They shared the first few layers to encode the state information. Additionally, Neunert et al.~\cite{neunert2020continuous} considered a hybrid policy that multiply the continuous and discrete actions distribution jointly, then optimized the policy with the maximum a posteriori policy optimization.

At present, there is limited research on RL for MIOC problems in engineering applications.
In unmanned aerial vehicles control, Samir et al.~\cite{samir2020age} used greedy-based and DDPG-based methods to update discrete scheduling decision and continuous traveling distance and direction, respectively.
Kamruzzaman et al.~\cite{kamruzzaman2021deep} adapted the architecture proposed by Fan et al.~\cite{fan2019hybrid} to a multi-agent framework for installation planning and controlling of shunts to enhance power system resilience.
Zhang et al.~\cite{zhang2020dynamic} and Ran et al.~\cite{ran2022optimizing} adopted a structure similar to P-DQN~\cite{xiong2018parametrized} for MIOC in mobile edge computing systems and data center scheduling and cooling systems, respectively. Nonetheless, they utilized the maximum Q-value as the target Q-value for the policy, as opposed to the sum of all Q-values in P-DQN. Their approach is referred to as deep deterministic actor-Q (DDAQ), and it serves as the foundation for our proposed TD3AQ algorithm in this paper.

For energy management in HEVs, the primary objective of the optimal control strategy is to determine the torque distribution that effectively balances fuel efficiency and driving performance~\cite{mei2023deep,tao2023terrain}.
Some research~\cite{li2019energy,wu2023application} utilized a similar framework to Hausknecht et al.~\cite{hausknecht2015deep} to obtain the discrete driving mode using the $\arg \max$ function.
Zhang et al.~\cite{zhang2020deep} treated the engine state (ON/OFF) as a binary classification problem and determined the discrete action output by a threshold.
Tang et al.~\cite{tang2021double} integrated DQN and DDPG techniques to separately derive values for throttle settings and gear ratios.
Similarly, Wang et al.~\cite{wang2022parameterized} employed an approach akin to that of P-DQN~\cite{xiong2018parametrized}, focusing on the regulation of engine torque and clutch states.
Gong et al.~\cite{gong2024plug} further extended the P-DQN algorithm by integrating a twin-delayed deep deterministic policy gradient (TD3)~\cite{fujimoto2018addressing} algorithm to better handle the hybrid action space involved.
These studies either simplify the problem to a single action space or consider the problem as a PAMDP. None of them solve the HEV energy management problem from a unified perspective.

Inspired by the aforementioned research, this paper proposes a novel hybrid-action reinforcement learning algorithm called twin delayed deep deterministic actor-Q (TD3AQ) for MIOC problems.
The algorithm is applied to address an energy management problem for a plug-in hybrid electric vehicle, aiming to demonstrate its effectiveness.
The main contributions of this study are summarized as follows:
\begin{itemize}
  \item A novel HARL algorithm, TD3AQ, is proposed to address the mixed-integer action space. TD3AQ combines the advantages of both TD3 and Q-learning methods, providing a unified solution for MIOC problems while stabilizing the training process..
  \item This study pioneers a comparative analysis of several prevalent RL algorithms within the domain of vehicular control applications, specifically focusing on PHEV energy management. Algorithms examined include DDAQ~\cite{ran2022optimizing}, PATD3~\cite{hausknecht2015deep}, and Rainbow~\cite{hessel2018rainbow} (discretized version). The evaluation results underscore TD3AQ's capability to achieve near-optimal solutions compared to dynamic programming, surpassing the performance of prior HARL baselines.
\end{itemize}

The structure of the remainder of this paper is outlined below.
Section~\ref{sec:harl} introduces the TD3AQ algorithm proposed in this work.
Section~\ref{sec:powertrain} provides an overview of the PHEV powertrain model and elucidates the energy management problem.
Section~\ref{sec:exp} presents the experimental results and offers insightful discussions.
Finally, in Section~\ref{sec:conclusion}, the paper is concluded, and avenues for future research are discussed.

\section{Twin Delayed Deep Deterministic Actor-Q}\label{sec:harl}
\subsection{Reinforcement Learning}
Reinforcement learning is a learning approach for optimal control by interacting with the environment.
The objective is to develop an agent that can learn to make optimal decisions by maximizing a cumulative reward signal.
Typically, this problem is formulated as a Markov decision process (MDP), which consists of a state space $\mathcal{S}$, an action space $\mathcal{A}$, an initial state distribution $p_0$, and a transition probability $p(s^{\prime}, r | s, a)$ that specifies the probability of transitioning to state $s^{\prime}$ and receiving reward $r$ when taking action $a$ in state $s$, where $r(s, a) \in \mathbb{R}$.
The agent's goal is to maximize the expected cumulative reward $\mathbb{E}\left[\sum_{t=0}^{\infty} \gamma^t r_t\right]$, where $\gamma$ is a discount factor.

According to the Bellman optimality principle, the value of a state under an optimal policy is the maximum expected return achievable from that state.
Reinforcement learning algorithms gradually refine value function estimates and converge to the optimal policy, which maximizes cumulative reward.

To achieve this goal, there are two mainstream algorithms: value-based methods and policy-based methods. Value-based methods learn the value function $V(s)$ or the action-value function $Q(s, a)$, which represent the expected cumulative reward starting from state $s$ or taking action $a$ in state $s$, respectively.
The optimal policy $\pi^*(s)$ can be derived from the value function $V(s)$ or the action-value function $Q(s, a)$ by selecting the action that maximizes $Q(s, a)$, i.e., $\pi^*(s) = \arg \max_a Q(s, a)$.
Policy-based methods, on the other hand, learn the parameterized policy $\pi_\theta(s)$ that can select actions without consulting a value function.
The policy is updated by gradient ascent on a performance measure $J(\theta)$, where $\theta$ represents the parameters of the policy network.
The policy network is a function approximator that maps the state $s$ to the action $a$.

Actor-critic methods combine value-based and policy-based approaches.
They learn both an approximate value function and a parameterized policy simultaneously, enabling efficient learning in complex environments.

\subsection{Mixed-Integer Action Space}

In this study, we explore the mixed-integer action space from the perspective of a parallel architecture, where discrete and continuous actions are independent of each other.
For instance, in a PHEV energy management problem, the gear shift command could be the discrete action, while the engine torque could be the continuous action.
These actions can be executed simultaneously and separately.
Such mixed-integer action spaces can be described by
\begin{equation}\label{eq:mixed_integer_action_space}
  \mathcal{A} = \{a = (a^d, a^c) \mid a^d \in \mathcal{A}_d, a^c \in \mathcal{A}_c \}.
\end{equation}
In this formulation, $\mathcal{A}_d$, a subset of $\mathbb{N}^k$, constitutes the discrete action spaces, while $\mathcal{A}_c$, a subset of $\mathbb{R}^n$, comprises the continuous action spaces. Here, $a^d$ and $a^c$ respectively signify the actions within these discrete and continuous domains.
The dimensionality of the action space is determined by $k + n$.
To handle the action space for multi-dimensional discrete variables, a composite action vector can be employed.
This vector has a length of $d_1\times d_2\times \cdots \times d_k$, where $d_i$ represents the number of discrete actions for the $i$-th dimension.
Each element of the vector corresponds to a distinct combination of discrete actions.
\subsection{TD3AQ Algorithm}
\begin{figure}[!t]
  \centering \includegraphics[width=0.32\textwidth]{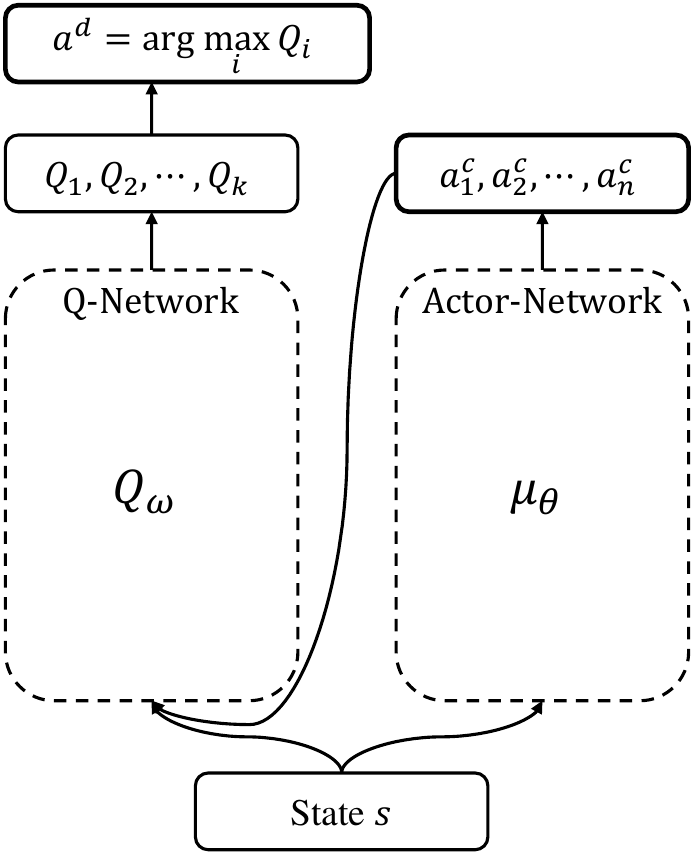}
  \caption{The forward propagation process of TD3AQ.
    The continuous action $a^c$ is generated by the actor network $\mu_\theta$, the subscript of $a^c$ represents the dimension of the action space, and the discrete action $a^d$ is selected by $\arg \max$ operation on the Q-values.}
  \label{fig:TD3AQ}
\end{figure}

To address the above mixed-integer action space, we propose a novel model-free algorithm, TD3AQ, which is based on the actor-critic framework.
The forward propagation process of TD3AQ is shown in Fig.~\ref{fig:TD3AQ}.
The actor network $\mu_\theta$ is used to generate the continuous actions $a^c$, then $a^c$ and state $s$ are fed into the Q-network to obtain the Q-values $Q_1, Q_2, \ldots, Q_k$ for each discrete action.
Finally, the discrete action with the highest Q-value is selected as $a^d$.
The details of the TD3AQ algorithm are elaborated below.
We first introduce the DDAQ algorithm \cite{zhang2020dynamic,ran2022optimizing}, which is a variant of DDPG that can handle the mixed-integer action space, and then we present the TD3AQ algorithm, which combines the DDAQ algorithm with the TD3 algorithm~\cite{fujimoto2018addressing}.
\subsubsection{Deep Deterministic Actor-Q}

Considering the vanilla DDPG algorithm~\cite{lillicrap2015continuous}, the Bellman equation is given by
\begin{equation}\label{eq:bellman_equation}
  Q_\omega(s, a) = \mathbb{E}_{r, s'} \left[ r + \gamma Q_{\omega} \left(s', \mu_{\theta}(s')\right) \right]
\end{equation}
where $\omega$ represents the parameters of the Q-network, $\mu_\theta$ represents the actor network that approximates the optimal action in the next state $s'$, and $\gamma$ is the discount factor.
The loss function to satisfy the Bellman equation is defined as
\begin{equation}\label{eq:bellman_loss}
  L(\omega) = \mathbb{E}_{(s, a, r, s') \sim \mathcal{B}} \left[ \left(y - Q_\omega(s, a)\right)^2 \right]
\end{equation}
where $\mathcal{B}$ is the replay buffer, a minibatch is sampled from $\mathcal{B}$ to update the Q-network, $y$ is the target value, which is defined as
\begin{equation}\label{eq:target_value}
  y = r + \gamma Q_{\omega'} \left(s', \mu_{\theta'}(s')\right)
\end{equation}
where $\omega'$ and $\theta'$ are the target networks to stabilize the training process. It is worth noting that the Eq.~\eqref{eq:bellman_equation} holds only when the actor network $\mu_\theta$ can provide the action that maximizes $Q_\omega(s, a)$. Assuming $Q_\omega$ is differentiable with respect to $a$, the actor network can be updated using the gradient ascent method by maximizing
\begin{equation}
  J(\theta) = \mathbb{E}_{s \sim \mathcal{B}} \left[ Q_\omega(s, \mu_\theta(s)) \right]
\end{equation}
Note that the parameters of $Q_\omega$ are fixed during the actor network update.

DDAQ adopts the same architecture as DDPG, but instead of approximating a single Q-value, it approximates $k$ Q-values for the $k$ discrete actions. Therefore, the Bellman equation can be rewritten as
\begin{align}\label{eq:bellman_equation_ddaq}
  Q_\omega(s, a) & = \mathbb{E}_{r, s'} \left[ r + \gamma \max_{a^d \in \mathcal{A}_d} Q_{\omega} \left(s', a^d, a^c\right) \right]                             \\
                 & = \mathbb{E}_{r, s'} \left[ r + \gamma \max_{a^d \in \mathcal{A}_d} \left( Q_{\omega} \left(s', a^d, \mu_{\theta}(s')\right) \right) \right]
\end{align}
Additionally, the loss function can be defined as
\begin{equation}\label{eq:bellman_loss_ddaq}
  L(\omega) = \mathbb{E}_{(s, a, r, s') \sim \mathcal{B}} \left[ \left(y - Q_\omega(s, a^d, a^c)\right)^2 \right]
\end{equation}
where
\begin{equation}\label{eq:target_value_ddaq}
  y = r + \gamma \max_{a^d \in \mathcal{A}_d} Q_{\omega'} \left(s', a^d, \mu_{\theta'}(s')\right)
\end{equation}

\begin{figure}[!t]
  \centering \includegraphics[width=0.28\textwidth]{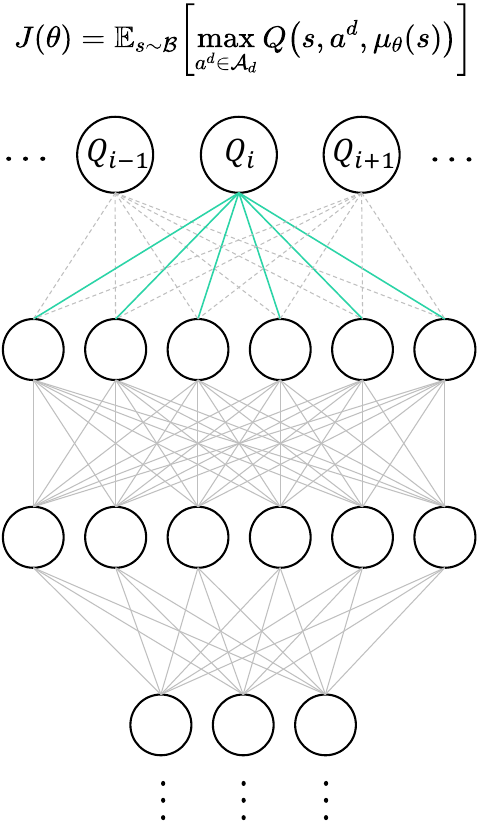}
  \caption{The back propagation process of DDAQ.
    Assuming that the $Q_i$ has the highest Q-value, DDAQ only updates the parameters pertaining to action $a^d=i$ (the solid green lines in the figure).}
  \label{fig:DDAQ_back}
\end{figure}

During the update process of the actor network, the objective function is defined based on the selection of the discrete action that exhibits the highest Q-value:
\begin{equation}\label{eq:objective_function_ddaq}
  J(\theta) = \mathbb{E}_{s \sim \mathcal{B}} \left[ \max_{a^d \in \mathcal{A}_d} Q\left(s, a^d, \mu_{\theta}(s)\right) \right]
\end{equation}
where the maximum Q-value is used as the objective function.
In previous work (P-DQN)~\cite{xiong2018parametrized}, the objective function was defined as the sum of all Q-values, which posed a problem when back propagating gradients: updating the weights in the Q-network during back propagation affected all weights in the network, including those unrelated to the action with the highest Q-value, potentially causing the algorithm to update in an incorrect direction.
DDAQ addresses this issue by using the maximum Q-value as the objective function, ensuring that only the weights corresponding to the action $a^d$ with the highest Q-value are updated.
The back propagation process of DDAQ is illustrated in Fig.~\ref{fig:DDAQ_back}.

\subsubsection{TD3AQ}

DDPG has shown potential in achieving good performance, but its robustness is limited due to its sensitivity to hyperparameters.
This is mainly because DDPG tends to overestimate Q-values, which results in poor policy updates.
To address this issue, we combine DDAQ and TD3 to form TD3AQ.
Three key modifications to improve the performance are introduced.

% \paragraph{Target policy smoothing}
1) Target policy smoothing:
A noise sampled from a normal distribution is added to the action generated by the target actor network. This technique acts as a regularizer that prevents the actor network from converging to a suboptimal solution. Specifically, the action was smoothed by adding a clipped noise, defined as follows:
\begin{equation}
  \tilde{a^c} = \operatorname{clip} (\mu_{\theta'}(s') + noise, a_{min}^c, a_{max}^c)
\end{equation}
where $a_{min}^c$ and $a_{max}^c$ are the lower and upper bounds of the continuous action, respectively. The \textit{noise} is defined as:
\begin{equation}
  noise = \operatorname{clip}(\epsilon, \epsilon_l, \epsilon_h), \quad \epsilon \sim \mathcal{N}(0, \sigma^2)
\end{equation}
where $\epsilon_l$ and $\epsilon_h$ are the lower and upper bounds of the noise $\epsilon$, respectively. The noise $\epsilon$ is sampled from a normal distribution with mean 0 and standard deviation $\sigma$.

2) Clipped double Q-learning:
To mitigate the overestimation of Q-values, a twin Q-network is introduced to DDAQ. When updating the Q-network, the target value is set as the minimum of the two chosen Q-values:
\begin{equation}
  y = r + \gamma \min_{j=1,2} \left(\max_{a^d \in \mathcal{A}_d} Q_{\omega_j'} \left(s', a^d, \tilde{a^c}\right)\right)
\end{equation}
then both Q-networks are updated by minimizing the mean squared error between the target value and the Q-value:
\begin{equation}
  L(\omega_j) = \mathbb{E}_{(s, a, r, s') \sim \mathcal{B}} \left[ \left(y - Q_{\omega_j}(s, a^d, a^c)\right)^2 \right], \quad j=1,2
\end{equation}

3) Delayed updates:
The actor network and target networks are updated less frequently than the Q-networks. This helps to stabilize the training process since slowing down the actor allows the critic to learn more accurate Q-values before letting it guide the actor. The networks are updated every two iterations as recommended in~\cite{fujimoto2018addressing}.

The proposed TD3AQ pseudo-code is shown in Algorithm~\ref{alg:td3aq}, with all the proposed improvements combined.

\begin{algorithm*}[!t]
  \caption{Pseudo-code of the TD3AQ Algorithm}
  \label{alg:td3aq}
  \LinesNumbered
  \SetKwFunction{Mod}{mod}
  \KwIn{
    Exploration noise $\eta = \mathcal{N}(0, \delta^2)$,
    exploration parameter $\varepsilon$,
    minibatch size $B$,
    empty replay buffer $\mathcal{B}$,
    target smooth noise $\epsilon \sim \mathcal{N}(0, \sigma^2)$,
    target smooth noise bounds $\epsilon_l$, $\epsilon_h$,
    discount factor $\gamma$,
    policy update frequency $K$,
    learning rates $\{\alpha, \beta, \rho\} \geq 0$.
  }

  Set up Q-network $Q_{\omega_1}$, $Q_{\omega_2}$, and actor network $\mu_\theta$ with initial random weights $\omega_1$, $\omega_2$, $\theta$\;

  Set up target networks $\omega_1^\prime \leftarrow \omega_1$, $\omega_2^\prime \leftarrow \omega_2$, $\theta^\prime \leftarrow \theta$\;

  Set up replay buffer $\mathcal{B}$\;

  \For{$t = 1, 2, \ldots, T$}{
  Clear local gradients $\mathrm{d}\omega_1 \leftarrow 0$, $\mathrm{d}\omega_2 \leftarrow 0$, $\mathrm{d}\theta \leftarrow 0$\;

  Compute continuous action with exploration noise $a_t^c = \operatorname{clip} \left(\mu_{\theta}\left(s_t\right) + \eta, a_{min}^c, a_{max}^c\right)$\;

  Compute discrete action $a_t^d$ according to $\varepsilon$-greedy
  \begin{equation*}
    a_{t}^d = \begin{cases}
      \text {a sample randomly from } \mathcal{A}_d,                                & \quad \text{with probability } \varepsilon   \\
      \arg \max_{a^d \in \mathcal{A}_d} Q_{\omega_1}\left(s_{t}, a^d, a_t^c\right), & \quad \text{with probability } 1-\varepsilon
    \end{cases}
  \end{equation*}

  Take action $a_t = (a_t^d, a_t^c)$, capture the obtained reward $r_t$, and transition to the next state $s_{t+1}$\;

  Record the transition $(s_t, a_t, r_t, s_{t+1})$ within buffer $\mathcal{B}$\;

  Draw a subset of transitions $\{(s_i, a_i, r_i, s_{i+1})\}_{i\in [B]}$ from $\mathcal{B}$\;

  Calculate the target continuous action
  \begin{equation*}
    \tilde{a_i^c} = \operatorname{clip}(\mu_{\theta'}(s_{i+1}) + \operatorname{clip}(\epsilon, \epsilon_l, \epsilon_h), a_{min}^c, a_{max}^c)
  \end{equation*}

  Calculate the target value
  \begin{equation*}
    y_i = r_i + \gamma \min_{j=1,2}\left(\max_{a^d \in \mathcal{A}_d} Q_{\omega_j^\prime}\left(s_{i+1}, a^d, \tilde{a_i^c}\right)\right)
  \end{equation*}

  Perform a single step of gradient descent to update the two Q-networks
  \begin{equation*}
    \omega_j \leftarrow \omega_j - \alpha \nabla_{\omega_j} \frac{1}{|B|} \sum_{i \in [B]} \left(Q_{\omega_j}\left(s_i, a_i^d, a_i^c\right) - y_i\right)^2, \qquad \text {for } j=1,2
  \end{equation*}

  \If{t \Mod $K$ == 0}{
    Perform a single step of gradient ascent to update the actor network
    \begin{equation*}
      \theta \leftarrow \theta + \beta \nabla_{\theta} \frac{1}{|B|} \sum_{i \in [B]} \max_{a^d} Q_{\omega_1}\left(s_i, a^d, \mu_\theta(s_i)\right)
    \end{equation*}

    Adjust the target networks as follows
    \begin{align*}
      \omega_j^\prime & \leftarrow \rho \omega_j + (1-\rho) \omega_j^\prime, \qquad \text {for } j=1,2 \\
      \theta^\prime   & \leftarrow \rho \theta + (1-\rho) \theta^\prime
    \end{align*}
  }
  }
\end{algorithm*}

\section{PHEV Energy Management Problem Formulation}\label{sec:powertrain}
To validate the proposed TD3AQ algorithm, this work considers a PHEV energy management problem, which presents a challenging MIOC problem as it requires simultaneous control of continuous power distribution and discrete gear selection for a highly non-linear hybrid system.
The section begins with an introduction to the PHEV powertrain modeling, followed by the MIOC problem formulation.

\subsection{PHEV Powertrain Configuration}

\begin{figure}[!t]
  \centering
  \includegraphics[width=0.48\textwidth]{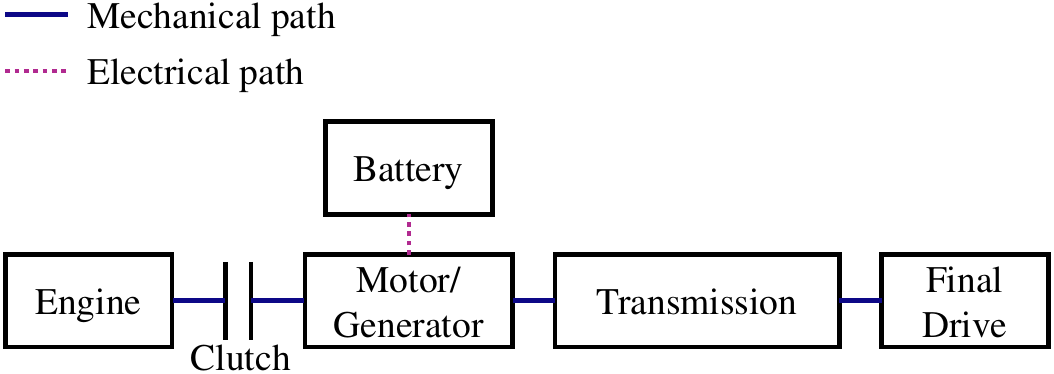}
  \caption{The parallel hybrid powertrain architecture.}
  \label{fig:powertrain}
\end{figure}

\begin{table}[!t]
  \caption{Technical parameter values of the PHEV.}
  \small
  \centering
  \begin{tabular}{lll}
    \toprule
    Symbol         & Parameters                      & Values          \\
    \midrule
    $\omega_{max}$ & Maximum shaft angular velocity  & 250 rad/s       \\
    $T_{b,max}$    & Maximum mechanical brake torque & 6000 Nm         \\
    $SOC_0$        & SOC initial value               & 0.9             \\
    $SOC_{l}$      & Minimum SOC                     & 0.3             \\
    $SOC_{h}$      & Maximum SOC                     & 0.9             \\
    $i_d$          & Final drive gear ratio          & 4.11            \\
    $\eta_f$       & Final drive efficiency          & 0.931           \\
    $\eta_g$       & Transmission efficiency         & 0.931           \\
    $Q_{max}$      & Battery capacity                & 26 Ah           \\
    $N_b$          & Number of battery cells         & 112             \\
    $\mu$          & Rolling resistance coefficient  & 0.01            \\
    $\rho$         & Air density                     & 1.1985 kg/m$^3$ \\
    $C_d$          & Air drag coefficient            & 0.65            \\
    $\theta$       & Road grade                      & 0 rad           \\
    $A$            & Vehicle cross-sectional area    & 6.73 m$^2$      \\
    $m$            & Vehicle mass                    & 5000 kg         \\
    $r$            & Tyre radius                     & 0.5715 m        \\
    $\Delta t$     & Control interval                & 1 s             \\
    \bottomrule
  \end{tabular}
  \label{tab:parameter_values}
\end{table}

This study focuses on a parallel hybrid electric vehicle~\cite{yu2020mixed}.
The powertrain architecture comprises an an electric motor, internal combustion engine, an automated manual transmission, and a clutch, illustrated in Fig.~\ref{fig:powertrain}.
The vehicle's HEV mode can be altered by engaging or disengaging the clutch between the engine and the motor, providing three modes of operation: motor-only, engine-only, and hybrid-drive.
The technical parameter values of the vehicle are listed in Table~\ref{tab:parameter_values}.

\subsection{Vehicle Dynamics}

The equations governing the vehicle dynamics are given as follows:
\begin{align}
   & T_w = \left[ m a_{cc} + \mu mg \cos(\theta) + mg \sin(\theta) + \frac{1}{2} A C_d \rho v^2 \right] r \\
   & \omega = \frac{v}{r} i_d i_g
\end{align}
Here, $T_w$ represents the torque at the wheels, $m$ denotes the mass of the vehicle, $a_{cc}$ is the acceleration, $\mu$ is the coefficient of rolling resistance, $g$ is the gravitational acceleration, $\theta$ describes the incline of the road, $A$ is the frontal cross-sectional area, $C_d$ is the coefficient of drag, $\rho$ stands for air density, $v$ indicates the vehicle's velocity, $r$ is the radius of the tires, $i_d$ refers to the final drive ratio, and $i_g$ represents the gear ratio of the transmission, selectable from a set ${\{6.25, 3.583, 2.22, 1.36, 1, 0.74\}}$.

The torque on the wheel can be denoted as:
\begin{align}\label{eq:tw}
   & T_w = (cT_{e,d}  + T_m)i_d i_g \eta_d \eta_g - T_b
\end{align}
where $T_{e,d}$ denotes the torque delivered by the engine, and $T_m$ is the torque from the motor, both contributing to the shaft torque.
$c$ is the clutch state with $c=1$ representing the clutch engaged and $c=0$ representing the clutch disengaged.
$\eta_d$ and $\eta_g$ represent the efficiencies of the final drive and transmission, respectively, and $T_b$ is the mechanical brake torque.

\subsection{Engine Model}

\begin{figure}[!t]
  \centering
  \includegraphics[width=0.4\textwidth]{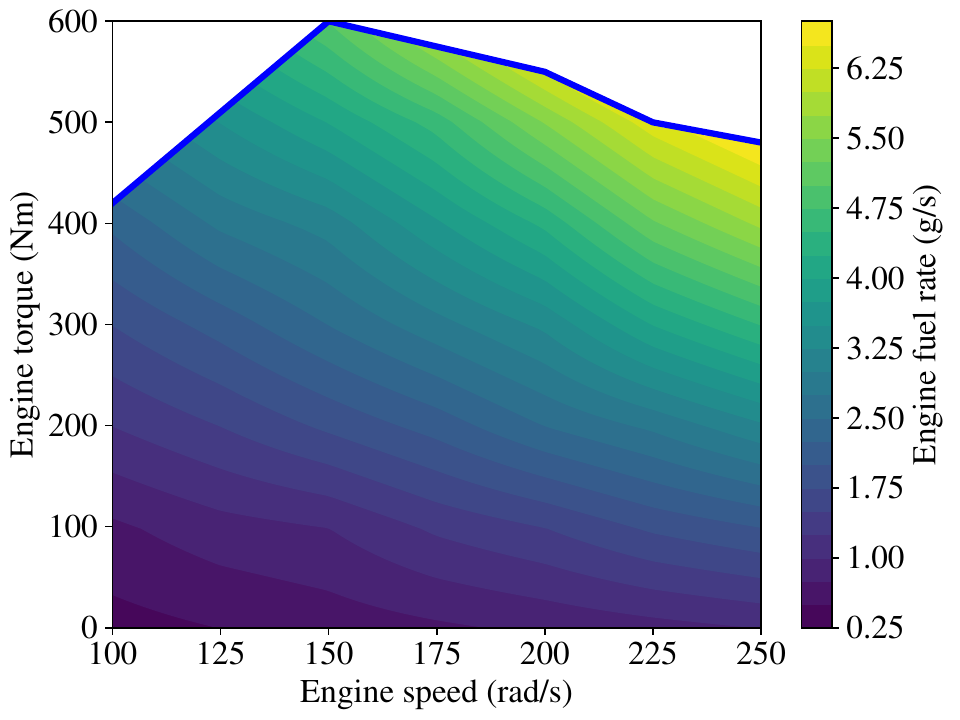}
  \caption{The engine fuel rate map, where the boundary represents the maximum engine torque.}
  \label{fig:mfdot_map}
\end{figure}

The engine torque consists of two distinct elements: the idle torque $T_{e,i}$ and the driving torque $T_{e,d}$, which can be expressed as:
\begin{align}
   & T_{e} = T_{e,i} + T_{e,d} \label{eq:Te_decomp}            \\
   & \text{s.t.} \quad 0 \leq T_e \leq T_{e,max} \label{eq:Te}
\end{align}
where $T_{e,max}= f_e(\omega_e)$ represents the maximum engine torque at a given angular velocity $\omega_e$, which is depicted in Fig.~\ref{fig:mfdot_map}.
When the clutch is disengaged, the torque from the engine, $T_{e,d}$, defaults to zero and $T_e$ shifts to $T_{e,i}$, which is primarily utilized to operate the auxiliary systems.
The operating speed of the engine is determined by the equation below:
\begin{align}
   & \omega _e=\left\{
  \begin{array}{l}
    \omega,\quad \text{if } c=1       \\
    \omega _{ei},\quad \text{if } c=0 \\
  \end{array}
  \right.                                                               \\
   & \text{s.t.} \quad \omega_{e} \ge \omega_{ei}, \quad \text{if } c=1
\end{align}
where $\omega_e$ represents the engine's angular velocity, and $\omega_{ei}$ specifies the engine's idle speed.
The constraint on $\omega_e$ is imposed to ensure that the engine speed does not drop below the idle speed when the clutch is engaged.

The fuel rate can then be described by:
\begin{equation}
  \dot{m}_f = f_f(\omega_{e}, T_e)
\end{equation}
where $\dot{m}_f$ is the fuel rate (kg/s), and $f_f$ is the fuel rate interpolation function, which is shown in Fig.~\ref{fig:mfdot_map}.

\subsection{Motor Model}

\begin{figure}[!t]
  \centering
  \includegraphics[width=0.4\textwidth]{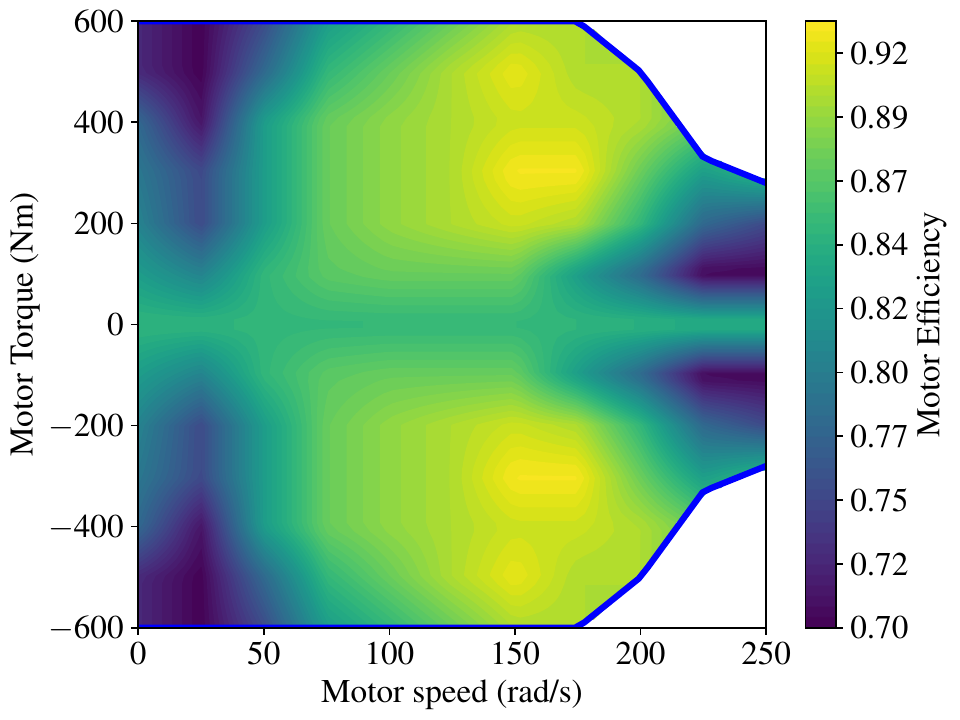}
  \caption{The motor efficiency map, where the upper and lower boundaries represent the maximum and minimum motor torque, respectively.}
  \label{fig:motor_efficiency}
\end{figure}

The motor torque $T_m$ is governed by the following constraints:
\begin{align}\label{eq:TmmaxTmmin}
  T_{m,max} & = f_m(\omega)                                    \\
  T_{m,min} & = -f_m(\omega)                                   \\
  T_{m,min} & \leq T_m \leq T_{m,max} \label{eq:Tm_constraint}
\end{align}
where $T_{m,max}$ and $T_{m,min}$ are the maximum and minimum motor torque at a given angular velocity, respectively. $f_m$ is an interpolation function that is used to calculate $T_{m,max}$ and $T_{m,min}$, which is depicted in Fig.~\ref{fig:motor_efficiency}.

\subsection{Battery Model}

\begin{figure}[!t]
  \centering
  \includegraphics[width=0.4\textwidth]{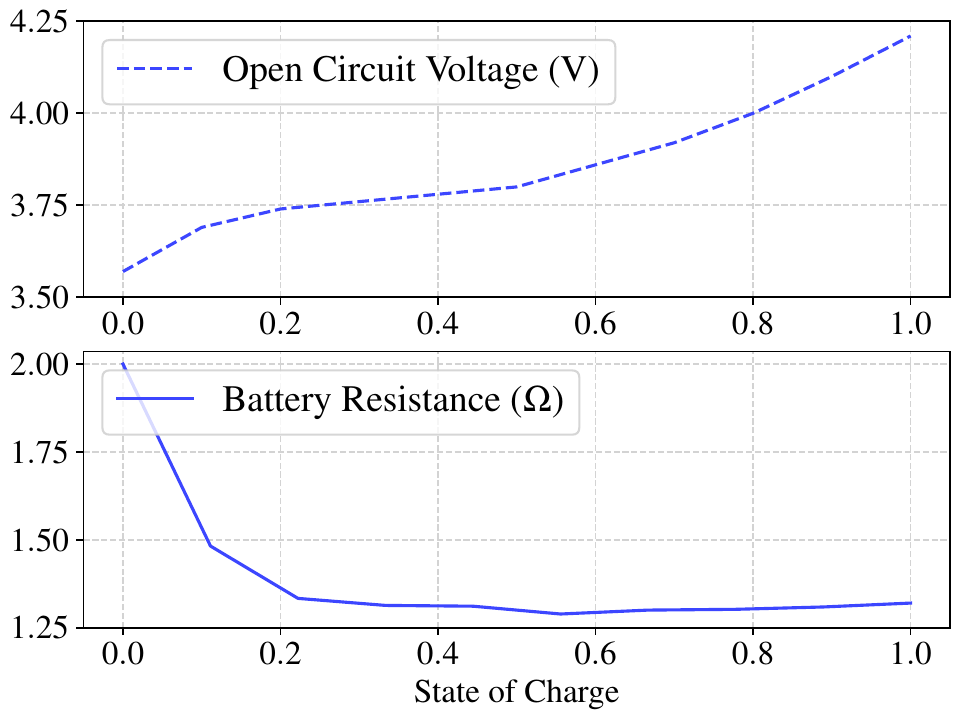}
  \caption{The battery cell's open-circuit voltage and resistance.}
  \label{fig:battery_parameters}
\end{figure}
We consider a lithium-ion battery pack consisting of 112 cells, where the cells are connected in series.
The battery model is based on the open-circuit voltage $E$ and state of charge (SOC). The SOC is calculated by:
\begin{align}\label{eq:soc}
  \dot{SOC} & = - \frac{I_b}{Q_{max}}                \\
  E I_b     & = I_b^2 R_b + P_b                      \\
  P_b       & = T_m \omega \eta_m^{\text{sgn}(-T_m)} \\
  \eta_m    & = f_\eta(\omega, T_m)
\end{align}
where $I_b$ is the net current, $Q_{max}$ is the battery capacity, $R_b$ is the battery resistance, $P_b$ is the battery power, $\eta_m$ is the motor efficiency, and $f_\eta$ is the motor efficiency interpolation function, which is shown in Fig.~\ref{fig:motor_efficiency}.
The battery open-circuit voltage and resistance are calculated by:
\begin{align}\label{eq:eoc}
  E   & = f_E(SOC)  \\
  R_b & = f_R(SOC).
\end{align}
where $f_E$ and $f_R$ are interpolation functions, as illustrated in Fig.~\ref{fig:battery_parameters}.
The battery SOC is constrained by:
\begin{align}\label{eq:soc_constraint}
  SOC_{l} \leq SOC \leq SOC_{h}
\end{align}
where $SOC_{l}$ and $SOC_{h}$ are the lower and upper SOC limits, respectively.

\subsection{Problem Formulation}

\begin{figure}[!t]
  \centering \includegraphics[width=0.48\textwidth]{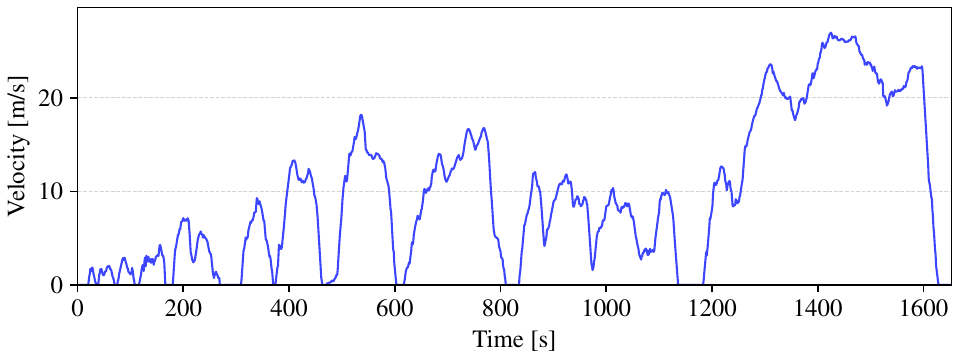}
  \caption{CHTC-LT drive cycle for light trucks.}
  \label{fig:chtc-lt}
\end{figure}
The primary objective of the PHEV energy management system is to develop a control strategy that minimizes the equivalent cost of fuel and electricity consumption, while simultaneously reducing the frequency of gear shifts and clutch engagement to enhance the overall driving experience.
The control inputs to the system are the engine torque, mechanical brake torque, gear ratio, and clutch state:
\begin{align}\label{eq:u}
  u & = [T_e, T_b, \varsigma, c]^\top
\end{align}
assuming that only a sequential shift pattern is considered, $\varsigma$ belongs to the set $\{-1, 0, 1\}$, where $\varsigma=-1$ represents downshift, $\varsigma=1$ represents upshift, and $\varsigma=0$ corresponds to sustainment.
The cost function to minimize can be expressed as:
\begin{align}\label{eq:J}
   & J = \sum_{T_0}^{T_{cyc}} (k_f \dot{m}_f + k_b P_b + p_{\varsigma} + p_{c}) \Delta t \\
   & p_{\varsigma}=\zeta \cdot \tilde{p},\quad \text{if } |gear_t-gear_{t-1}| > 0        \\
   & p_{c}=\xi \cdot \tilde{p},\quad \text{if } |c_t-c_{t-1}| > 0
\end{align}
where $k_f$ and $k_b$ are the fuel and electricity prices, respectively, $\tilde{p}$ denotes the reference penalty, established at the median value of fuel consumption, $p_{\varsigma}$ and $p_{c}$ are the penalties for gear shift and clutch engagement/disengagement, respectively, $\zeta$ and $\xi$ are the penalty coefficients, $\Delta t$ is the control interval, $T_0$ and $T_{cyc}$ are the initial and final time steps of the drive cycle, respectively. The drive cycle utilized for training is the China heavy-duty transient cycle for light trucks (CHTC-LT), depicted in Fig.~\ref{fig:chtc-lt}.

\section{Experiments and Discussion}\label{sec:exp}
\subsection{TD3AQ for PHEV Energy Management}
\subsubsection{State}
The observation space of the RL agent is defined as the state of the system, which is a vector of the following variables:
\begin{equation}\label{eq:state}
  s = [ v, a_{cc}, T_{w}, SOC, i_g, c]
\end{equation}
\subsubsection{Action}
The action space is defined by the torque relationship in Eq.~\eqref{eq:Te} as follows:
\begin{equation}\label{eq:action}
  a = [T_e, \varsigma, c]
\end{equation}
where $T_e$ is continuous, and $\varsigma$ and $c$ are discrete, making it an MIOC problem.
In negative demand torque, regenerative braking is prioritized for battery charging. If regenerative braking is insufficient, mechanical braking provides the remaining torque.
When the required torque is positive, the mechanical braking torque is zeroed out, with both the engine and motor contributing to meet the torque requirements.
Thus, the proposed action determines the vehicle's torque distribution entirely.

To handle the multi-discrete action space involving two actions, $\varsigma$ and $c$, we employ a composite action vector.
This vector, with a length of $3 \times 2 = 6$, represents specific combinations of three gear shift actions and two clutch actions.
This approach effectively manages the multi-discrete action space while maintaining a compact representation.
\subsubsection{Reward}
In addition to the cost function defined in Eq.~\eqref{eq:J}, it is imperative to ensure that the system dynamics remain within specific constraints.
Consequently, the reward function incorporates both fuel usage and penalties for constraint violations.
The detailed hand-crafted reward function is defined as follows:
\begin{align}\label{eq:reward}
   & r_{cost}    = - (k_f\dot{m}_f+k_bP_b+p_{\varsigma} + p_{c})                                                                                                \\
   & p_{\omega}  = \tilde{p}, \quad \text{if } \omega > \omega_{max}                                                                                            \\
   & p_{T_m}     = 10\tilde{p}, \quad \text{if } T_m > T_{m,max}                                                                                                \\
   & p_{SOC}     =\begin{cases}
                    0,                                                   & \text{if } SOC_{l} \leq SOC \leq SOC_{h} \\
                    10\tilde{p} \cdot \frac{|SOC - SOC_{h}|}{1-SOC_{h}}, & \text{if } SOC > SOC_{h}                 \\
                    10\tilde{p} \cdot \frac{|SOC - SOC_{l}|}{SOC_{l}},   & \text{if } SOC < SOC_{l}
                  \end{cases} \\
   & r         = r_{cost} - p_{\omega} - p_{T_m} - p_{SOC}
\end{align}
where $p_{\omega}$, $p_{T_m}$, and $p_{SOC}$ are the penalties for exceeding the constraints on the motor torque, shaft angular velocity, and SOC, respectively.
The penalty for exceeding the SOC constraint is designed as a linear function of the SOC deviation from the desired range, which extends from the lower bound $SOC_l$ to the upper bound $SOC_h$.
Therefore, the greater the deviation of the SOC from the desired range, the higher the penalty.

\subsubsection{Hyperparameters}
\begin{table}[!t]
  \caption{Hyperparameters of TD3AQ.}
  \small
  \begin{center}
    \begin{tabular}{ll}
      \toprule
      Parameter                       & Value                      \\
      \midrule
      Soft target update coefficient  & 0.001                      \\
      Reward discount factor          & 0.99                       \\
      Optimizer                       & Adam~\cite{kingma2014adam} \\
      Actor-network learning rate     & 0.0001                     \\
      Q-network learning rate         & 0.001                      \\
      Experience replay memory size   & 200\,000                   \\
      Minibatch size                  & 128                        \\
      Exploration noise               & $\mathcal{N}(0, 0.02)$     \\
      Actor-network hidden layer size & $64\times64$               \\
      Q-network hidden layer size     & $64\times64$               \\
      \bottomrule
    \end{tabular}
  \end{center}
  \label{tab:td3aq_para}
\end{table}
The hyperparameters of the proposed TD3AQ are summarized in Table~\ref{tab:td3aq_para}, both actor and Q networks consist of two hidden layers with 64 neurons each. The continuous action noise is modeled as a zero-mean Gaussian distribution with a standard deviation of 0.02.

\subsection{Comparison Benchmarks}
Four benchmark methods, DP, DDAQ~\cite{zhang2020dynamic}, PATD3~\cite{hausknecht2015deep}, and Rainbow~\cite{hessel2018rainbow}, are selected for comparison.
Although we implemented Hybrid PPO following the approach described by Neunert et al.~\cite{neunert2020continuous}, it was excluded from the comparison due to its inferior performance, incurring a 29.80\% higher cost compared to DP.

DP utilizes the MATLAB toolbox created by Sundstr\"om et al.~\cite{sundstroem2009generic}. 
The continuous engine torque is segmented at intervals of 25~Nm.
Empirical experiments have demonstrated that finer discretization does not improve the performance significantly (less than 0.05\% improvement in cost with 10 Nm discretization while requiring seven times more computation time).
Since Rainbow is designed for discrete action spaces, it adopts the same discretization scheme as DP.
All RL algorithms utilize Optuna~\cite{akiba2019optuna} to tune the hyperparameters.

\subsection{Experimental Setup}

The training process involves a total of 200\,000 timesteps, where each timestep represents one second of driving time.
Model evaluation was performed every one episode, with each episode corresponding to one drive cycle.
The final model for testing was selected based on the highest return achieved during evaluation.
% All experiments were conducted on a desktop computer equipped with an AMD Ryzen 9 5950X CPU and an NVIDIA GeForce RTX 3080 Ti GPU\@.
The source code of the algorithm is available at \url{https://github.com/xujinming01/TD3AQ}.

\subsection{Learning Ability Assessment}

\begin{figure}[!t]
  \centering
  \includegraphics[width=0.48\textwidth]{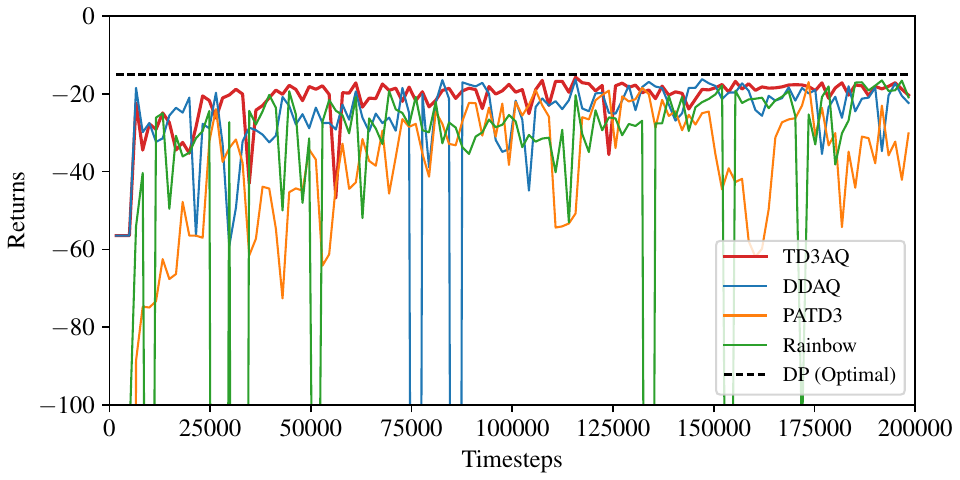}
  \caption{The reward curves during training.}
  \label{fig:reward_curve}
\end{figure}

The reward curves during training are shown in Fig.~\ref{fig:reward_curve}.
It can be observed that, with the exception of PATD3, all algorithms start to converge after 50,000 timesteps.
Furthermore, TD3AQ exhibits relatively lower fluctuations after convergence.
In contrast, DDAQ and Rainbow encounter notable drops in rewards intermittently.
Taken together, these results highlight TD3AQ's expedited convergence and enhanced stability throughout the training process.

\subsection{Optimal Control Performance}

% \begin{figure*}[!t]
%   \centering
%   \subfigure[Engine torque]{
%     \includegraphics[width=0.44\textwidth]{fig-engine_torque.pdf}
%   }
%   \subfigure[Motor torque]{
%     \includegraphics[width=0.44\textwidth]{fig-motor_torque.pdf}
%   }
%   \\
%   \subfigure[Gear shift sequence]{
%     \includegraphics[width=0.44\textwidth]{fig-ig_comparison.pdf}
%   }
%   \subfigure[Clutch state (0 = disengaged, 1 = engaged)]{
%     \includegraphics[width=0.44\textwidth]{fig-clutch_comparison.pdf}
%   }
%   \caption{The hybrid action output of DP and TD3AQ for CHTC-LT.}
%   \label{fig:hybrid_action}
% \end{figure*}

\begin{figure*}[!t]
  \centering
  \subfigure[Engine torque]{
    \includegraphics[width=0.95\textwidth]{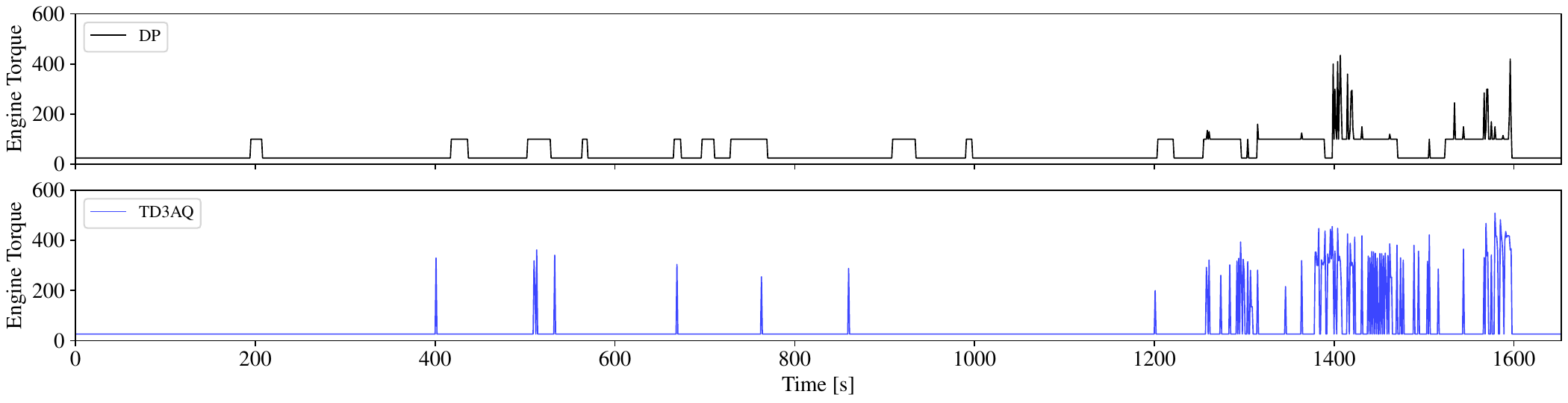}
  }
  \\
  \subfigure[Motor torque]{
    \includegraphics[width=0.95\textwidth]{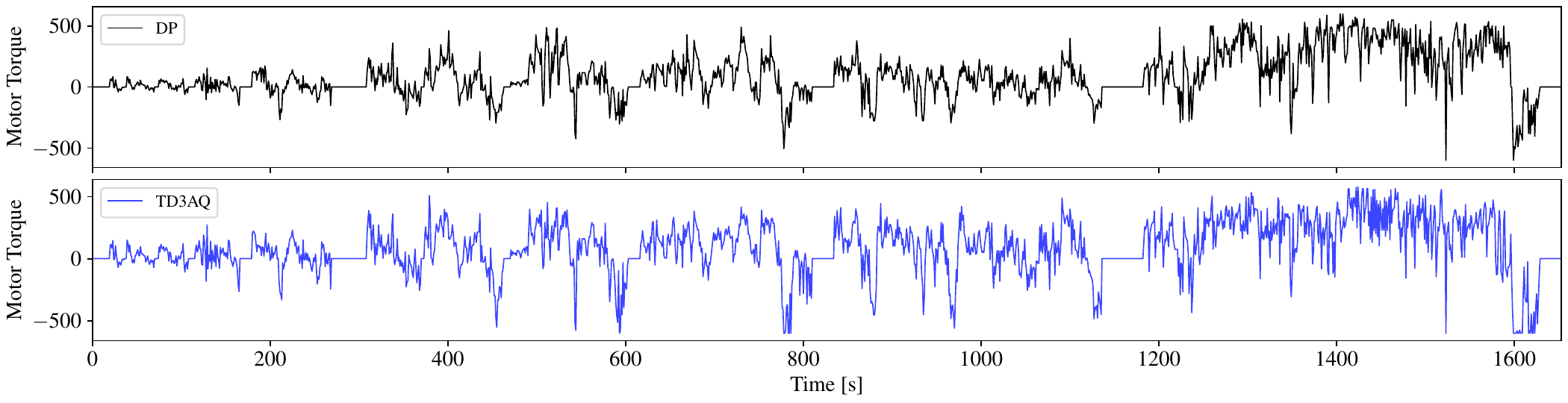}
  }
  \\
  \subfigure[Gear shift sequence]{
    \includegraphics[width=0.95\textwidth]{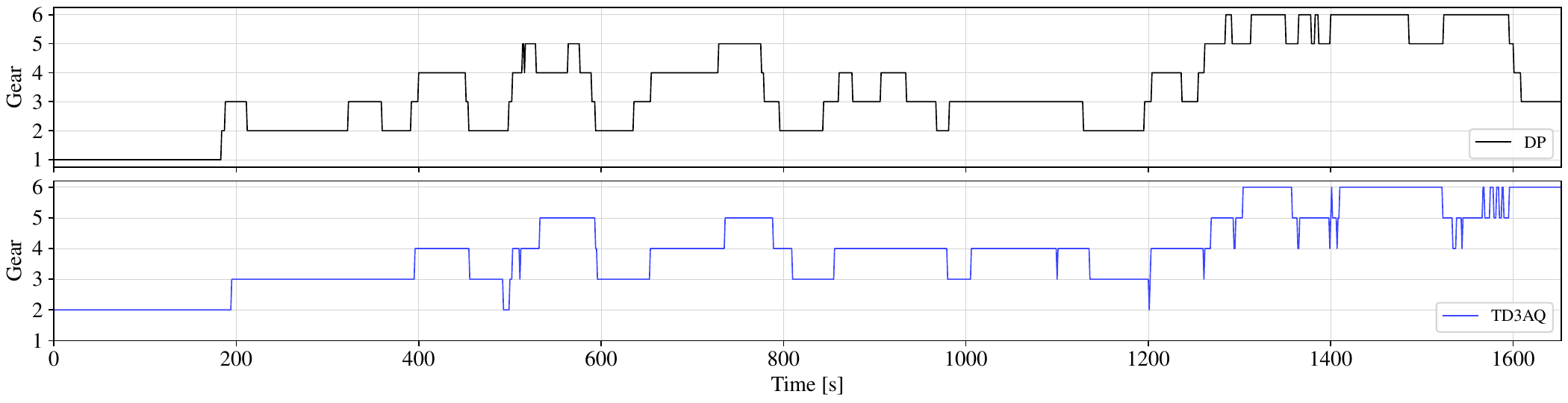}
  }
  \\
  \subfigure[Clutch status (1 for engaged, 0 for disengaged)]{
    \includegraphics[width=0.95\textwidth]{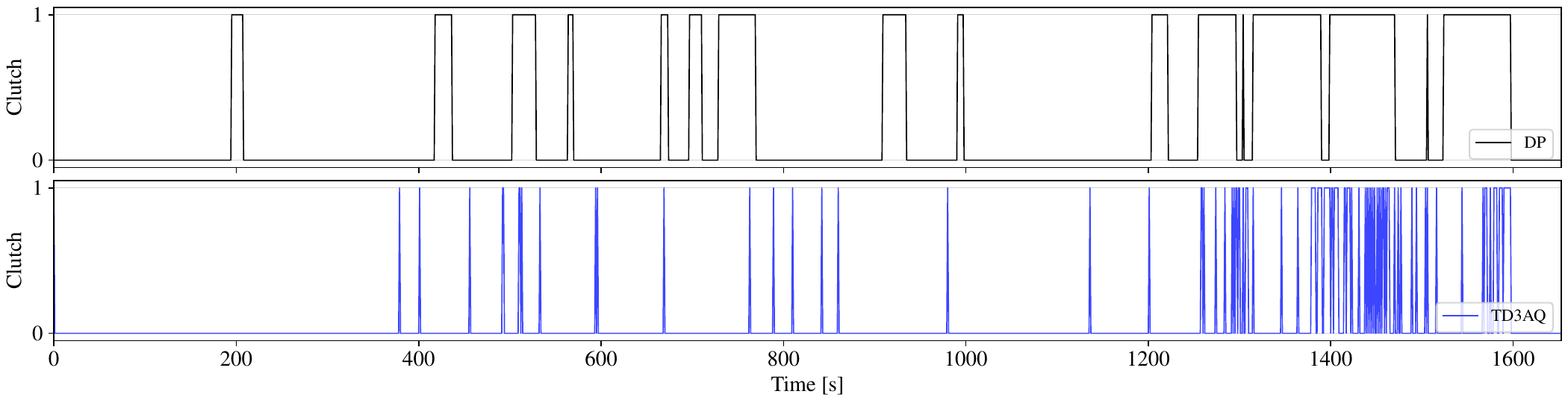}
  }
  \caption{The hybrid action results from TD3AQ and DP for CHTC-LT.}
  \label{fig:hybrid_action}
\end{figure*}

\begin{table}
  \caption{Results when trained and tested on the same drive cycle (CHTC-LT).}
  \small
  \centering
  \begin{tabular}{lrrrrrrrr}
    \toprule
    Strategy & $\tau_{sh}$ & $\tau_{cl}$ & $\nu_{T_m}$ & $\nu_\omega$ & $\nu_{soc}$ & Cost/\textyen & Gap    \\
    \midrule
    DP       & 52          & 32          & -           & -            & -           & 15.04         & -      \\
    TD3AQ    & 55          & 136         & 0           & 0            & 0           & 15.74         & 4.69\% \\
    DDAQ     & 126         & 183         & 0           & 0            & 0           & 16.25         & 8.06\% \\
    PATD3    & 70          & 106         & 1           & 1            & 0           & 16.49         & 9.66\% \\
    Rainbow  & 171         & 160         & 1           & 0            & 0           & 16.12         & 7.19\% \\
    \bottomrule
  \end{tabular}
  \label{table:fuel_CHTC_LT}
\end{table}

Fig.~\ref{fig:hybrid_action} displays the hybrid action output of DP and TD3AQ employed in CHTC-LT. It is evident that TD3AQ exhibits similar output trends to DP, however, with a higher frequency of gear shifts and clutch state changes.
Notably, DP operates under the assumption that the entire drive cycle is known in advance.
In contrast, RL algorithms only possess access to the current state, yielding suboptimal solutions.
Table~\ref{table:fuel_CHTC_LT} presents the simulation results for CHTC-LT,
where $\tau_{sh}$ and $\tau_{cl}$ denote the number of gear shifts and clutch engagements, respectively, $\nu_{T_m}$, $\nu_\omega$, and $\nu_{soc}$ denote the instances where motor torque, shaft angular velocity, and SOC exceed the predefined constraints.
Among the RL algorithms, TD3AQ exhibits the lowest cost, with a gap of 4.69\% compared to DP. Furthermore, TD3AQ and DDAQ do not violate any constraints, while PATD3 and Rainbow demonstrate a few violations in terms of motor torque and shaft angular velocity constraints.
\begin{figure*}[!t]
  \centering \includegraphics[width=0.95\textwidth]{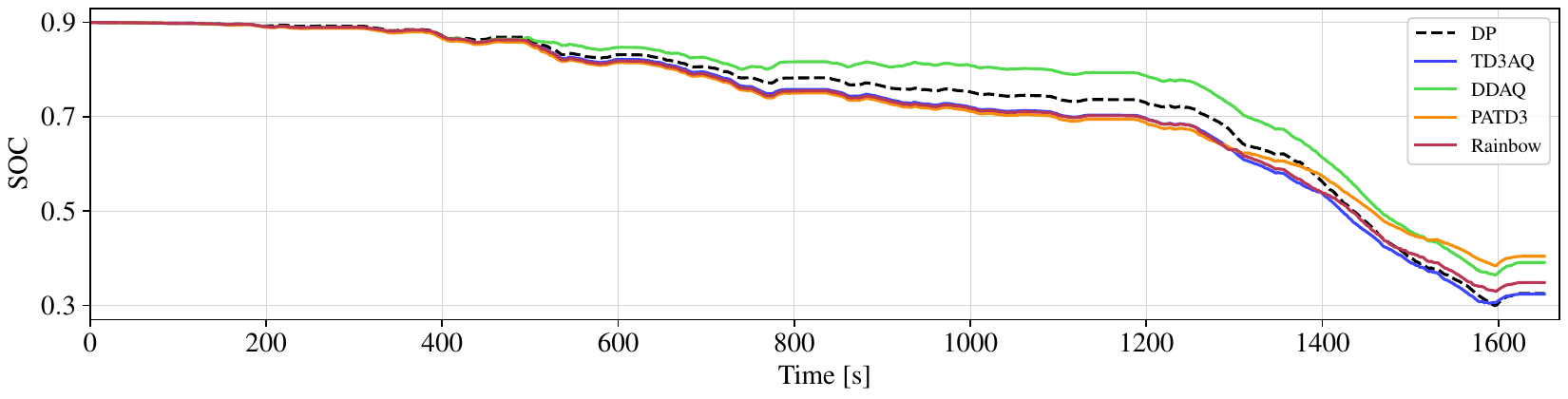}
  \caption{SOC trajectories of CHTC-LT drive cycle.}
  \label{fig:soc-CHTC-LT}
\end{figure*}

Fig.~\ref{fig:soc-CHTC-LT} depicts the SOC trajectories for the RL algorithms and DP.
It is observed that the RL algorithms generally exhibit similar SOC trajectories to DP, except during the final portions of the drive cycle.
Given that the lower bound $SOC_l$ is set to 0.3, all the algorithms reach an SOC of 0.3 by the end of the drive cycle, with the exception of DDAQ and PATD3.
Importantly, it is worth emphasizing that there are no violations of the SOC constraint, indicating that RL algorithms can effectively learn to regulate the SOC within the desired range without prior knowledge of the specific drive cycle.

\begin{table}[!t]
  \centering
  \caption{Wall-clock time comparison for CHTC-LT.}
  \small
  \begin{tabular}{lrr}
    \toprule
    Strategy & Training time & Execution time \\
    \midrule
    TD3AQ    & 0.21 h        & 0.46 ms        \\
    DDAQ     & 0.20 h        & 0.45 ms        \\
    PATD3    & 0.27 h        & 1.04 ms        \\
    Rainbow  & 0.85 h        & 1.15 ms        \\
    DP       & N/A           & 96599 ms       \\
    \bottomrule
  \end{tabular}
  \label{tab:time_CHTC_LT}
\end{table}

Table~\ref{tab:time_CHTC_LT} presents the time consumption comparison for CHTC-LT, where execution time represents the time required to execute a single action during testing.
While DP can obtain the optimal solution for each step in a single calculation, it comes at the expense of a significant amount of time, with its single solution time being six orders of magnitude greater than RL\@.
This substantial discrepancy in time consumption highlights the efficiency of the RL algorithms, as indicated by their step times falling within the millisecond range.
Consequently, RL demonstrates promising potential for real-time optimal control applications.

\subsection{Generalization Results}

\begin{table}[!t]
  \centering
  \caption{Generalization results for new drive cycles.}
  \small
  \begin{tabular}{llrrrrrr}
    \toprule
    Test cycle               & Strategy & $\nu_{T_m}$ & $\nu_\omega$ & $\nu_{soc}$ & Cost/\textyen & Gap     \\
    \midrule
    \multirow{5}{*}{JE05}    & DP       & -           & -            & -           & 13.60         & -       \\
                             & TD3AQ    & 0           & 0            & 0           & 14.09         & 3.60\%  \\
                             & DDAQ     & 0           & 0            & 0           & 15.15         & 11.40\% \\
                             & PATD3    & 0           & 0            & 0           & 16.68         & 22.60\% \\
                             & Rainbow  & 0           & 0            & 0           & 14.59         & 7.21\%  \\
    \midrule
    \multirow{5}{*}{HD-UDDS} & DP       & -           & -            & -           & 9.27          & -       \\
                             & TD3AQ    & 6           & 10           & 0           & 10.12         & 8.55\%  \\
                             & DDAQ     & 22          & 0            & 0           & 10.28         & 10.20\% \\
                             & PATD3    & 15          & 125          & 0           & 10.81         & 15.88\% \\
                             & Rainbow  & 24          & 5            & 0           & 10.05         & 7.73\%  \\
    \midrule
    \multirow{5}{*}{WHVC}    & DP       & -           & -            & -           & 5.51          & -       \\
                             & TD3AQ    & 0           & 0            & 0           & 5.72          & 3.64\%  \\
                             & DDAQ     & 0           & 1            & 0           & 6.70          & 21.33\% \\
                             & PATD3    & 0           & 8            & 0           & 5.74          & 3.95\%  \\
                             & Rainbow  & 1           & 0            & 0           & 5.94          & 7.69\%  \\
    \bottomrule
  \end{tabular}
  \label{tab:generalization}
\end{table}

To evaluate the generalization capabilities of the proposed HARL algorithm TD3AQ, the model trained exclusively on the CHTC-LT drive cycle is tested on three additional drive cycles: JE05, HD-UDDS, and WHVC. The simulation results are summarized in Table~\ref{tab:generalization}.

The findings demonstrate that the TD3AQ algorithm exhibits commendable generalization performance when applied to new drive cycles.
In comparison with other RL algorithms, TD3AQ closely approximates the cost achieved by the DP strategy, with an average gap of 5.26\% across the three cycles.
However, the remaining RL algorithms display a larger deviation in terms of cost.

Furthermore, it is worth noting that all RL algorithms exhibit a higher frequency of constraint violations in the new drive cycles, particularly PATD3 and Rainbow.
This issue primarily arises from the inherent limitation of RL in enforcing hard constraints during the training process.

\section{Conclusion}\label{sec:conclusion}
In this paper, we present a novel reinforcement learning algorithm called TD3AQ for solving mixed-integer optimal control problems.
The TD3AQ algorithm uses actor-Q networks to directly generate continuous and discrete actions from the current state.
We evaluate the proposed TD3AQ algorithm on a hybrid electric vehicle energy management problem and show that it achieves satisfactory performance with a cost gap of 4.69\% compared to the DP strategy.
TD3AQ's ability to generalize is demonstrated through the results on three new drive cycles.
The algorithm's sub-millisecond response time indicates its significant potential for real-time, mixed-integer optimal control applications.
However, note that the algorithm carries the risk of violating constraints, which, if unaddressed, could potentially lead to vehicle and battery damage in real-world scenarios.
Future work will focus on extending TD3AQ to more complex, higher-dimensional problems to further substantiate its adaptability and robustness, and developing HARL algorithms with safety guarantees.

\section*{Acknowledgment}
This research received partial funding from the Guangzhou Basic and Applied Basic Research Program under Grant 2023A04J1688, and was also supported by South China University of Technology faculty start-up fund.



\bibliographystyle{elsarticle-num}
\bibliography{ref}




% \end{thebibliography}
\end{document}